# An alternative scheme of angular-dispersion analyzers for high-resolution medium-energy inelastic X-ray scattering

Xian-Rong Huang*

*Advanced Photon Source, Argonne National Laboratory, Argonne, Illinois 60439, USA*
E-mail: xiahuang@aps.anl.gov

The development of medium-energy inelastic X-ray scattering (IXS) optics with meV and sub-meV resolution has attracted considerable efforts in recent years. Meanwhile, there are also concerns or debates about the fundamental and feasibility of the involved schemes. Here the central optical component, the back-reflection angular-dispersion monochromator or analyzer, is analyzed. The results show that the multiple-beam diffraction effect together with transmission-induced absorption can noticeably reduce the diffraction efficiency, although it may not be a fatal threat. In order to improve the efficiency, a simple four-bounce analyzer is proposed that completely avoids these two adverse effects. The new scheme is illustrated to be a feasible alternative approach for developing meV- to sub-meV-resolution IXS spectroscopy.

## 1. Introduction

Inelastic X-ray scattering (IXS) spectroscopy with energy resolution of ~1 milli-electronvolt (meV) is a powerful technique for studying vibrational dynamics in solids, liquids and biological materials (Burkel, 2000). In addition to the conventional high-resolution IXS spectroscopy that must be carried out with high-energy photons (typically $E > 20$ keV), meV- or even sub-meV-resolution IXS optics for medium energies around 10 keV have attracted considerable attention and development efforts recently. The advantages of the latter include higher photon flux generated from undulators and higher momentum resolution at medium energies. Most importantly, the new optics may make it possible to perform meV or sub-meV IXS experiments using a large number of emerging modern medium-energy synchrotron light sources and X-ray free-electron lasers that are usually unable to provide sufficient high-energy photon flux.

The top challenge of meV and sub-meV IXS optics for medium energies is that one cannot use the conventional back-reflection analyzers since the intrinsic back-reflection spectral bandwidths of silicon or germanium are one or two orders broader than 1 meV at $E \sim 10$ keV. To surmount this obstacle, Shvyd'ko *et al.* (2006, 2007) have proposed the concept of back-reflection angular-dispersion monochromators and analyzers. Instead of using the entire bandwidth, this scheme uses extremely asymmetric crystals to disperse the back-reflected X-rays along slightly different directions according to their wavelengths. Afterwards, a large-incidence-small-exit Bragg reflection with a narrow angular acceptance is used to angularly select a small portion of the dispersed spectrum to generate a sub-meV bandpass. Implementation of this concept has been aggressively pursued, particularly at the National Synchrotron Light Source II (NSLS-II) and the Advanced Photon Source (APS).

However, to date the experimental progress has been relatively slow, and full-scale monochromators and analyzers with convincing resolution and efficiency close to the theoretical values are yet to be demonstrated. Meanwhile, there are also concerns about the fundamentals or the practical feasibility of this concept. For example, it is unclear how the multiple-beam diffraction effect involved in the almost exact back-reflection geometry affects the diffraction efficiency. Daunting challenges also include fabrication of the meter-long dispersing crystals or the alternative 'comb crystals' (Shvyd'ko, 2008) with stringent requirements of lattice homogeneity and surface perfection, fabrication of collimating multilayer mirrors (Honnicke *et al.*, 2010), and crystal mounting and stability (Cai, 2010).

In this paper, we first present a detailed analysis of the multiple-beam diffraction effect involved in the back reflection to understand how it affects the efficiency of the back-reflection-based optics. Then we propose a new scheme that is based on near back reflections and can improve the efficiency by ~50% by completely removing multiple-beam diffraction and transmission-induced X-ray absorption.

## 2. Multiple-beam diffraction in CDTS

The simplest configuration of the back-reflection angular-dispersion optics (consisting of silicon crystals) proposed by Shvyd'ko *et al.* (2006, 2007) is shown in Fig. 1, where the monochromatization process can be described by the following four steps. (i) A polychromatic and divergent incident beam is first collimated by 220 Bragg reflection from the upper surface of the thin crystal *C* (*i.e.* the incidence divergence $\Delta\alpha$ is reduced to $|b_C|\Delta\alpha$ after *C*, $b_C$ the asymmetric factor of 220 reflection). (ii) The collimated beam is then back reflected by the dispersing crystal *D* with 008 Bragg reflection. Due to angular dispersion (Huang *et al.*, 2008), the back reflected beam becomes a dispersion fan, in which different wavelengths are diffracted along slightly different directions according to $K_{008x} = K_{0x} + h_{008x}$, or

$$\cos\theta_e = (\lambda/d)\cos\varphi - \cos\theta, \qquad (1)$$

where $\theta$ and $\theta_e$ are the incident and exit angles of crystal *D*, respectively, $\varphi$ is the offcut angle of *D*, $d$ is the spacing of (008) lattice planes, and $\lambda$ is the X-ray wavelength (see Fig. 2 for the definitions of $K_{008x}$, $K_{0x}$ and $h_{008x}$). (iii) The dispersion fan can selectively pass through crystal *C* by Borrmann anomalous transmission effect. (iv) Finally, the selector *S* with the large-incidence-small-exit 220 reflection only diffracts X-rays within a narrow angular range (~5 µrad) of the transmitted fan, resulting in a sub-meV bandpass. Therefore, this is a *C*ollimation-*D*ispersion-*T*ransmission-*S*election (CDTS) monochromatization process. In the following, we will assume that crystals *C* and *S* have the same asymmetric angle of 19° for 220 reflection (with Bragg angle $\theta_B = 20.7°$ at the 008 back-reflection energy of 9.1315 keV, and $b_C = -0.0465$ for the small-incidence geometry). The thickness of *C* is 0.2 mm.

Note that the Borrmann anomalous transmission effect uniquely utilized in the CDTS diffraction process occurs strongly only in the large-incidence-small-exit diffraction geometry (Kishino, 1974). The inset in Fig. 1 shows the enhanced transmission curve of crystal *C*, where the maximum transmissivity is only 0.75, indicating that transmission-induced absorption is considerable during this single step. More importantly, the incident angle corresponding to the transmission peak is smaller than that of the 220 Bragg reflection peak by 5 µrad. Consequently, crystal *S* must be tilted by 5 µrad with respect to *C* to select the transmission peak. The 5-µrad angular difference leads to a strict constraint that the effective diffraction angle of the 008 reflection from *D* must be exactly around 90° − 2.5 µrad, which is an almost *exact back reflection*. Under this condition, however, it is known that parasitic reflections can be activated (Sutter *et al.*, 2001), leading to multiple-beam diffraction that may reduce the 008 reflection efficiency, as illustrated below.

When exact 008 back reflection occurs, the Bragg conditions of 404, $\bar{4}04$, 044 and $0\bar{4}4$ reflections (with Bragg angles of 45°) are all satisfied (Nikulin *et al.*, 2003). For σ-polarization in Fig. 2 where the electric field **E** is parallel to **y**, 044 and $0\bar{4}4$ reflections are forbidden. Thus, we only need to treat the 4-beam coplanar diffraction geometry. For σ-polarization with $\nabla \cdot \mathbf{E} \equiv 0$, Maxwell's equations lead to $K^2 \varepsilon \mathbf{E} = -\nabla^2 \mathbf{E}$, of which the Fourier transformation form in the crystal can be written as

$$k_\mathbf{h}^2 E_\mathbf{h} = K^2 \sum_{\mathbf{h}'} \varepsilon_{\mathbf{h}-\mathbf{h}'} E_{\mathbf{h}'}, \tag{2}$$

where $K = 2\pi/\lambda$, $E_\mathbf{h}$ is the amplitude of the plane wave component along **y**, $\mathbf{k_h}$ is the internal diffracted wavevector associated with the reciprocal lattice vector **h**, and $\varepsilon_\mathbf{h}$ is the Fourier component of the permittivity $\varepsilon$ and is related to the crystal susceptibility component by $\varepsilon_0 = 1 + \chi_0$ or $\varepsilon_\mathbf{h} = \chi_\mathbf{h}$ for $\mathbf{h} \neq 0$. For convenience, we rescale all the wavevectors and reciprocal lattice vectors by a factor of $1/K$. In terms of Fig. 2, equation (2) then becomes

$$\begin{pmatrix} k_0^2 - \varepsilon_0 & -\varepsilon_{008} & -\varepsilon_{404} & -\varepsilon_{404} \\ -\varepsilon_{008} & k_{008}^2 - \varepsilon_0 & -\varepsilon_{404} & -\varepsilon_{404} \\ -\varepsilon_{404} & -\varepsilon_{404} & k_{404}^2 - \varepsilon_0 & -\varepsilon_{008} \\ -\varepsilon_{404} & -\varepsilon_{404} & -\varepsilon_{008} & k_{\bar{4}04}^2 - \varepsilon_0 \end{pmatrix} \begin{pmatrix} E_0 \\ E_{008} \\ E_{404} \\ E_{\bar{4}04} \end{pmatrix} = 0. \tag{3}$$

The incident wavevector above the crystal is $\mathbf{K}_0 = K_{0x}\mathbf{x} + K_{0z}\mathbf{z}$ with $K_{0x} = \cos\theta$ and $K_{0z} = -\sin\theta$. The forward refracted wavevector in the crystal can be written as $\mathbf{k}_0 = K_{0x}\mathbf{z} + q\mathbf{z}$, where $q$ is a complex quantity to be determined. Then the diffracted wavevector in the crystal can be written as $\mathbf{k}_h = \mathbf{k}_0 + \mathbf{h} = (K_{0x} + h_x)\mathbf{x} + (h_z + q)\mathbf{z}$ for $\mathbf{h} = \mathbf{0}$, 008, 404 or $\bar{4}04$, where $h_x$ and $h_z$ are the tangential and vertical components of **h**. Consequently, we have

$$\begin{aligned} k_0^2 &= K_{0x}^2 + q^2, \\ k_h^2 &= (K_{0x} + h_x)^2 + h_z^2 + 2h_z q + q^2. \end{aligned} \tag{4}$$

Based on equations (4), equation (3) can be written as an eigenvalue-eigenvector equation

$$(q^2 \mathbf{I} - q\mathbf{V} + \mathbf{B})\tilde{\mathbf{E}} = 0, \tag{5}$$

where $\tilde{\mathbf{E}} = (E_0, E_{008}, E_{404}, E_{\bar{4}04})^T$, **I** is the 4 × 4 identity matrix, **V** is a diagonal matrix with $V_{11} = 0$, $V_{22} = -2h_{008z}$, $V_{33} = -2h_{404z}$ and $V_{44} = -2h_{\bar{4}04z}$, and

$$\mathbf{B} = \begin{pmatrix} B_{11} & -\varepsilon_{008} & -\varepsilon_{404} & -\varepsilon_{404} \\ -\varepsilon_{008} & B_{22} & -\varepsilon_{404} & -\varepsilon_{404} \\ -\varepsilon_{404} & -\varepsilon_{404} & B_{33} & -\varepsilon_{008} \\ -\varepsilon_{404} & -\varepsilon_{404} & -\varepsilon_{008} & B_{44} \end{pmatrix} \tag{6}$$

with $B_{11} = K_{0x}^2 - \varepsilon_0$, $B_{22} = (K_{0x} + h_{008x})^2 + h_{008z}^2 - \varepsilon_0$, $B_{33} = (K_{0x} + h_{404x})^2 + h_{404z}^2 - \varepsilon_0$ and $B_{44} = (K_{0x} + h_{\bar{4}04x})^2 + h_{\bar{4}04z}^2 - \varepsilon_0$. Equation (5) now has the same form as equation (10) by Colella (1974) and one can use the same method to obtain eight eigenvalues of $q$. For thick crystals, only four eigenvalues with Im($q$) > 0 are valid. Based on the four corresponding eigenvectors, one may use the boundary conditions (*i.e.* the continuity of the tangential electric and magnetic fields across the surface) to obtain the 008 and 404 reflectivity. This method is rigorous even for extremely grazing geometry (Huang & Dudley, 2003; Cho *et al.*, 2004). The calculations in this paper are based on the susceptibility components $\chi_0 = -(11.745 + i0.21741) \times 10^{-6}$, $\chi_{008} = -(2.8139 + i0.16846) \times 10^{-6}$, $\chi_{404} = -(4.5461 + i0.19137) \times 10^{-6}$ and $\chi_{220} = -(7.1968 + i0.21058) \times 10^{-6}$.

Since the 008 reflection from crystal *D* in CDTS is a nearly exact back reflection, we only need to consider photon energies within (or very close to) the 008 exact back-reflection bandwidth, which is between $\Delta E_1$ = 40 meV and $\Delta E_2$ = 67 meV relative to the Bragg energy $E_B = hc/(2d)$ (*h* the Plank constant and *c* the velocity of light in free space). Here the shift of the bandwidth from $E_B$ toward the higher energy range is owing to the slight X-ray refraction effect. With the offcut angle set to $\varphi$ = 2°, Fig. 3 shows the single-crystal angular Darwin curves of 008 reflection at four different photon energies, where for comparison, the 2-beam Darwin curves calculated with 404 and $\bar{4}04$ reflections artificially ignored are also presented. Overall, the 4- and 2-beam Darwin curves coincide with each other for most incident angles and photon energies, indicating that 008 reflection overwhelmingly dominates the 4-beam diffraction process.

In the central ranges of Fig. 3, however, the multiple-beam effect does appear, which leads to the small 008 reflectivity dips that can be seen more clearly in Figs. 3(*a*′)-3(*d*′). As mentioned above, in the CDTS multi-crystal diffraction process, only the back-reflected X-rays angularly deviated from the opposite direction of the incident beam by ~5 μrad are effective. The dashed lines in Figs. 3(*a*′)-3(*d*′) correspond to this condition (*i.e.* $\theta_e - \theta$ = 5 μrad in Fig. 2). Therefore, although the parasitic 404 and $\bar{4}04$ reflections are not significant over a wide range in Figs. 3(*a*)-3(*d*), unfortunately the small-range multiple-beam diffraction conditions almost exactly overlap the stringent CDTS multi-crystal diffraction conditions in Figs. 3(*a*′)-3(*d*′). In other words, the effective CDTS diffraction process always involves parasitic 404 and $\bar{4}04$ reflections of crystal *D*.

Next, we incorporate the above 4-beam diffraction principles into the dynamical theory computations of the full CDTS diffraction process. First let us simulate the *D* crystal rocking curve since this is the simplest way to demonstrate the angular dispersion principle in experiments (Shvyd'ko, 2006). Note that rocking crystal *D* (with crystals *C* and *S* fixed) is equivalent to rocking the dispersion fan in Fig. 1. Then the fixed selector *S* continuously selects different wavelengths from the rotating fan. The relative photon energy $\Delta E$ selected by *S* linearly increases with $\Delta\theta_D$, where $\Delta\theta_D$ is the rocking angle of crystal *D* relative to the incident angle $\theta = \varphi$ (the same as $\Delta\theta$ in Fig. 3). For example, the $\Delta\theta_D$ positions corresponding to the four photon energies in Fig. 3 are marked in Fig. 4(*a*), from which one may understand the two rocking curves in Fig. 4(*a*) more clearly based on Fig. 3.

The 4-beam rocking curve in Fig. 4(*a*) was calculated under a white beam with a flat spectrum and with vertical divergence of $\Delta\alpha$ = 80 μrad incident on crystal *C*. The output intensity *I* after crystal *S* was convoluted with both the incidence divergence and the photon energies, *i.e.*

$$I(\Delta\theta_D) = \int_{\Delta\alpha}\int_E R_C R_D T_C R_S d\alpha \, dE, \tag{7}$$

where $R_C$, $R_D$ and $R_S$ are the wavelength- and incident-direction-dependent reflectivity functions of crystals $C$, $D$ and $S$, respectively, and $T_C$ is the transmission function of crystal $C$. As a reference, the 2-beam rocking curve calculated with $R_D$ in equation (7) replaced by the 2-beam 008 reflectivity (with $40\bar{4}$ and $\bar{4}04$ reflections artificially ignored) is also shown in Fig. 4(*a*). Obviously, the multiple-beam diffraction effect reduces the CDTS output in all the strong diffraction range. Near the center of the rocking curves, the intensity involving 4-beam diffraction drops from the corresponding intensity of the 2-beam diffraction by 16%. Fortunately, the peak intensity difference between the two curves is only 6.3%.

When CDTS is used as an analyzer in IXS experiments, the angle of crystal $D$ will be fixed, preferably at the position slightly higher than the angle of the $I(\Delta\theta_D)$ peak for efficiency and stability reasons. The reflectivity curve as a function of the photon energy (*i.e.* energy resolution function) shown in the inset in Fig. 4(*a*) was calculated with crystal $D$ fixed at $\Delta\theta_D$ = 128 μrad, from which one can see that the peak of this integrated reflectivity curve, $R(\Delta E) = \int_{\Delta\alpha} R_C R_D T_C R_S d\alpha$, is $R_{max}^{CDTS} = 0.37$ with an energy resolution (*i.e.* bandwidth) of $\Delta E_{CDTS}$ = 0.66 meV. Here $R_{max}^{CDTS}$ is almost the maximum efficiency of CDTS for the current crystal parameters.

Recently, Shvyd'ko (2008) also proposed a variant of the CDTS scheme (also see Shvyd'ko *et al.*, 2011), which can be described as a five-step *C*ollimation-*D*ispersion-*T*ransmission-*D*ispersion-*S*election (CDTDS) setup shown in the inset in Fig. 4(*b*). In this variant, an extra dispersing crystal ($D_2$) is added to enhance the angular dispersion so that the bandwidth can nominally be reduced by half compared with the CDTS setup for the same asymmetric angles. However, since the X-rays undergo 008 back reflection twice in CDTDS, the efficiency loss caused by 4-beam diffraction becomes worse. Detailed calculations in Fig. 4(*b*) show that for the CDTDS configuration with $\varphi$ = 2° (for both of the dispersing crystals), the relative peak intensity loss is 12% and the intensity loss near the rocking curve center is 25%. Meanwhile, the maximum reflectivity for $\Delta\alpha$ = 80 μrad is $R_{max}^{CDTDS} = 0.34$ and the energy resolution is $\Delta E_{CDTDS}$ = 0.44 meV [see Fig. 5(*c*), not exactly half of $\Delta E_{CDTS}$ owing to the Borrmann effect]. Note that the exact shape of the 4-beam (rather than the 2-beam) rocking curve in Fig. 4(b) has been experimentally verified recently at APS (Shvyd'ko *et al.*, 2011).

The bandwidth of CDTDS (as well as CDTS) is nearly proportional to $\varphi$. However, the multiple-beam diffraction effect also increases with $\varphi$. For example, for achieving ~1 meV resolution with CDTDS, $\varphi$ can be increased to 5° (corresponding to shorter $D$ crystals). Unfortunately, at $\varphi$ = 5°, the relative peak intensity loss caused by multiple-beam diffraction is about 24% (with $R_{max}^{CDTDS} = 0.31$ and $\Delta E_{CDTDS}$ = 1.1 meV), indicating that CDTDS is not well suited for meV-resolution optics.

From the above calculations, we may draw the following conclusions. (i) In principle, the novel CDTS and CDTDS optics indeed are capable of achieving sub-meV resolution (but with daunting technical challenges). (ii) The multi-crystal diffraction process always involves noticeable multiple-beam diffraction in the 008 back reflection. The efficiency loss caused by this effect, however, is not fatal although it is unfavorable. (iii) The maximum theoretical peak reflectivity is only about 0.37 for CDTS and 0.34 for CDTDS based on the above typical crystal parameters, which is not as high as previously expected (in comparison with conventional monochromators) except that the angular acceptance here is much broad (~ 0.1 mrad). (iv) In addition to the multiple-beam diffraction effect, the other factor that attributes to the relatively low reflectivity is X-ray absorption caused by the Borrmann transmission process through the thin crystal $C$.

Note that the above calculations are based on perfect Si crystals. If the crystals are not sufficiently perfect, two undesirable situations may arise. First, Borrmann anomalous transmission is extremely sensitive to lattice strains and defects, *i.e.* the latter can easily reduce or even completely destroy the anomalous transmission. [Due to this reason, Borrmann transmission X-ray topography has been used as a high-sensitivity technique for imaging crystal defects (Kishino, 1974).] Here the thin *C* crystal is required to be only ~0.2 mm thick with both surfaces well polished and free of strains or defects. Fabrication of such high-perfection crystals is difficult. Strains caused by crystal mounting can also affect the efficiency as well as the resolution. Experiments indeed have shown that crystal *C* is a troublesome component in CDTS and CDTDS.

Second, similar to the 220 reflection from *C*, the small-incidence-large-exit 404 reflection alone is also a strong reflection with broad bandwidth and angular acceptance (*i.e.* its Bragg condition is quite loose). The reason why 404 reflection is not significant in CDTS and CDTDS (consisting of perfect crystals) is that it is suppressed by the multiple-beam diffraction mechanism. However, if the suppression is broken by surface imperfections (*e.g.* strains and roughness on the *D* surface), it is possible that the 404 reflection may become a major alternative route for the X-rays to be 90° diffracted instead of being back-reflected. This could remarkably reduce the 008 reflection efficiency. In the following, we will propose a modified scheme to completely avoid these two adverse situations.

## 3. Four-bounce CDDS scheme

The straightforward way to avoid multiple-beam diffraction is to move the 008 Bragg angle away from 90°, which can be realized by the simple four-bounce *C*ollimation-*D*ispersion-*D*ispersion-*S*election (4B-CDDS) scheme in Fig. 5(*a*). This scheme also removes the troublesome thin crystal and the associated Borrmann transmission process. Here first note that X-ray angular dispersion occurs for any asymmetric reflection ($\varphi \neq 0$). According to equation (1), the *dispersion rate* is

$$\frac{\Delta \theta_e}{\Delta \lambda} = -\frac{\cos \varphi}{d \sin \theta_e}, \qquad (8)$$

which represents how fast the exit direction $\theta_e$ varies with $\lambda$ for a collimated polychromatic incident beam (*i.e.* the incident angle $\theta$ is constant). To achieve a sufficiently high dispersion rate, one only needs to make both $\theta_e$ and $\varphi$ sufficiently small, and the exact back reflection geometry in Fig. 2 is not absolutely necessary.

In Fig. 5(*a*), since crystal *C* is the same as that in Fig. 1, the angular acceptance of 220 reflection is still ~0.1 mrad. The 008 reflection of the first dispersing crystal $D_1$ has an asymmetric factor $b_{D1}$ with $|b_{D1}| > 1$, which results in the following situation. Consider a *monochromatic* wave component with slight divergence $\Delta \alpha_{D1}$ incident on $D_1$. After 008 reflection from $D_1$, the divergence $\Delta \alpha_{D1}$ is *magnified* to $|b_{D1}|\Delta \alpha_{D1}$. Now consider another monochromatic and divergent component but with a different wavelength. Its divergence is also magnified by $D_1$. Unfortunately, the two divergent components after $D_1$ may overlap. The overlapped region corresponds to the situation that the two different wavelengths are diffracted along the same direction, which degrades the dispersion quality. This is different from the back-reflection geometry in CDTS and CDTDS where the asymmetric factor is always −1. However, if we set the asymmetric factor of the second dispersing crystal $D_2$ in Fig. 5(*a*) to be $b_{D2} = 1/b_{D1}$, the divergence variation can be completely cancelled out. Subsequently, the polychromatic beam diffracted from $D_2$ becomes a regular dispersion fan with different wavelengths along slightly different directions without overlap (the same as that in CDTDS). From this fan the selector $S$ can select a narrow bandpass.

In Fig. 5(*a*), since the four crystals are all independent thick crystals, one has more freedom to choose their parameters as desired (in addition to the convenience of fine surface polishing). For example, one may choose different reflections for $S$ except that its angular acceptance must be the same as the beam divergence $\Delta\alpha_{D1}$ after the collimator $C$. Overall, the major working mechanisms of 4B-CDDS and CDTDS are very similar.

As an example, let us set the 008 Bragg angles of $D_1$ and $D_2$ to be 89° in Fig. 5(*a*), corresponding to a Bragg energy of 9.1330 keV. Then the 008 reflection becomes a pure two-beam diffraction case with 404 and $\bar{4}04$ reflections completely vanishing. If $D_1$ and $D_2$ have the same offcut angle $\varphi = 2°$, the incident and exit angles of $D_1$ are $\theta_1 \approx 3°$ and $\theta_{e1} \approx 1°$, respectively ($b_{D1} \approx -3$). Accordingly, the incident and exit angles of $D_2$ are $\theta_2 \approx 1°$ and $\theta_{e2} \approx 3°$, respectively ($b_{D2} \approx -1/3$). For a collimated polychromatic beam incident on crystal $D_1$, it can be proved that the combined dispersion rate after $D_2$ is

$$\frac{\Delta\theta_{e2}}{\Delta\lambda} = -\frac{\cos\varphi}{d}\left(\frac{1}{\sin\theta_{e1}} + \frac{1}{\sin\theta_{e2}}\right). \tag{9}$$

For the CDTDS configuration in Fig. 4(b) with $\theta_{e1} \approx \theta_{e2} \approx \varphi$, equation (9) becomes

$$\frac{\Delta\theta_{e2}}{\Delta\lambda} = -\frac{2\cos\varphi}{d\sin\varphi}, \tag{10}$$

which is the combined dispersion rate of $D_1$ and $D_2$ for CDTDS and is twice that of equation (8) for a single back-reflection dispersing crystal. From equations (9) and (10) one can find that the ratio between the combined dispersion rate of 4B-CDDS with $\varphi = 2°$, $\theta_{e1} \approx 1°$ and $\theta_{e2} \approx 3°$ and that of CDTDS with $\theta_{e1} \approx \theta_{e2} \approx \varphi = 2°$ is 1.3, which indicates that here 4B-CDDS is slightly more dispersive than the above CDTDS configuration (*i.e.* the energy resolution of the current 4B-CDDS setup should be slightly higher).

Now we choose the reflection of $S$ in Fig. 5(*a*) to be 440 (or 224) for the purpose of making the output beam direction far away from the forward direction to avoid possible background on the detector. We again assume that the incident beam has initial vertical divergence of $\Delta\alpha = 80$ μrad, which is reduced to $\Delta\alpha_{D1} = 3.7$ μrad after the collimator $C$. Under this condition, the asymmetric factor of $S$ with 440 reflection only needs to be $b_S = -6.7$ for an angular acceptance of 3.7 μrad. Based on these parameters, we calculated the resolution function of 4B-CDDS in Fig. 5(*c*) under the optimized diffraction conditions. The energy resolution is $\Delta E_{\text{CDDS}} = 0.52$ meV, which is slightly worse than $\Delta E_{\text{CDTDS}} = 044$ meV in Fig. 4(*b*). According to equations (9) and (10), however, we have predicted $\Delta E_{\text{CDDS}} < \Delta E_{\text{CDTDS}}$. This discrepancy is caused by the selective Borrmann transmission process through the thin crystal of CDTDS, which is equivalent to an extra filtering process.

However, the maximum reflectivity of 4B-CDDS in Fig. 5(*c*) is $R_{\max}^{\text{CDDS}} = 0.53$, about 50% higher than those of CDTS and CDTDS (0.37 and 0.34, respectively, see §2), which is a significant improvement for flux-hungry IXS. The efficiency gain here is obviously owing to the removal of multiple-beam diffraction and X-ray absorption caused by the Borrmann transmission process. In terms of the diffraction efficiency, therefore, 4B-CDDS is superior to CDTS and CDTDS.

Also plotted in Fig. 5(*c*) is the resolution function of the CDTDS setup in Fig. 4(*b*) with the resolution of $\Delta E_{\text{CDTDS}} = 0.44$ meV, from which one can see one of the most distinct features of CDTDS, the extremely steep tails (Shvyd'ko *et al.*, 2011) resulting from Borrmann transmission through the thin crystal (at the cost of absorption). (For CDTS, only one side is very steep.) Since the bare 4B-CDDS does not use the Borrmann effect, this feature is largely lost. Nevertheless, the 4B-CDDS multi-crystal diffraction analyzer (similar to the 4-bounce monochromator in Fig. 6) has a much better resolution function (*i.e.* higher spectral contrast) than the conventional single-bounce back-reflection analyzer, of which the resolution function is close to the Lorentzian distribution in Fig. 5(*c*). The resolution function of 4B-CDDS can be greatly improved, which will be our future studies. On the other hand, the peak sharpness of CDTDS could be easily destroyed by diffuse scattering from crystal surface roughness (particularly for comb crystals).

One of the major challenges of CDTS and CDTDS is that the dispersing crystals must be extremely long (while the lattice homogeneity must be controlled at the $< 10^{-7}$ level). If we assume that the height of the incident beam is 1 mm before *C*, the footprint of the beam on the *D* crystals is 620 mm long for CDTS and CDTDS with $\varphi = 2°$. But for the current 4B-CDDS, the footprint becomes 410 mm because of the larger glancing angles $\theta_1 = \theta_{e2} = 3°$. To further shorten the *D* crystal length, one can also adopt the 'comb crystals' [see Fig. 5(*b*)] proposed by Shvyd'ko (2008) for CDTS and CDTDS. Compared with CDTDS, the comb crystals for 4B-CDDS require less thin crystal plates since the total footprint is shorter. Meanwhile, the gap between the plates is larger owing to the larger glancing angles $\theta_1$ and $\theta_{e2}$ (which makes the fabrication more feasible and easier).

As another example, if we relax the energy resolution to 1 meV, the required length of $D_1$ and $D_2$ is only about 200 mm for 4B-CDDS (with $\theta_1 = 6°$ and $\theta_{e1} = 2°$). Under this condition, comb crystals may be unnecessary. Therefore, the 4B-CDDS analyzer can be designed for both meV and sub-meV resolution. By contrast, the CDTS and CDTDS can only work in the sub-meV range (requiring comb crystals) since the efficiency loss resulting from multiple-beam diffraction becomes worse with increasing $\varphi$.

In addition, the reason why 008 back reflection has been (carefully) chosen for CDTS and CDTDS is that this reflection has the minimum multiple-beam diffraction effect (although it still exists) compared with other back reflections (Shvyd'ko, 2007). For the 4B-CDDS scheme that is free of multiple-beam diffraction, this restriction is removed such that one can freely choose any desirable back reflections (energies) to implement the analyzer.

A technical constraint of 4B-CDDS is that the three crystals must be well separated in order to make the beam reflected from $D_1$ bypass *C*. For example, if the incident beam before *C* is 1 mm high and the incident and exit angles on $D_1$ are 3° and 1°, respectively, the distance between *C* and the top of $D_1$ must be greater than 0.6 m. The distance between *S* and the bottom of $D_2$ has the same requirement. (For meV analyzers, the distances can be shorter.) CDTDS does not have this restriction and can be more compact if comb crystals are used.

Based on the 4B-CDDS analyzer, we thus propose the 'hybrid' meV or sub-meV IXS spectroscopy beamline sketched in Fig. 6. Note that 4B-CDDS cannot be used as the monochromator as it changes the beam direction. Here we propose the use of the in-line 4-bounce high-resolution monochromator (4B-HRM) (Yabashi *et al.*, 2001). Obviously, the 4B-HRM can be described as a collimation-collimation-dispersion-selection monochromator, *i.e.* it is also based on the angular dispersion mechanism. However, since it does not use back reflections, the 4B-HRM can be designed to work at any energies. For example, at APS this design has been successfully implemented at $E \sim 9.4$ keV with resolution of 1 meV and efficiency of 36% (close to the theoretical values) achieved from Krypton nuclear resonance analyses (Toellner, 2008). Therefore, the 4B-HRM is a mature monochromator for achieving meV to sub-meV resolution and has been routinely used in various applications. Another advantage of the 4B-HRM is that, unlike CDTDS, it does not require temperature scan to scan the photon energy. Instead, energy scan can be realized by crystal rotations. Also note that CDTDS causes more significant virtual source spread than the 4B-HRM, which makes microfocusing of the monochromatized beam extremely difficult (Huang *et al.*, 2011). Furthermore, in CDTDS, high-energy harmonics are directly transmitted through the thin crystal along the forward direction without diffraction, which is a severe disadvantage in experiments.

In Fig. 6 we may simply choose 642 reflection (Bragg angle $\theta_B$ = 69.3° for $E$ = 9.1330 keV) for all the four crystals of the 4B-HRM with asymmetric factors $b_1 = b_2 = 1/b_3 = 1/b_4 = -0.081$. Under the condition that the divergence of the incident undulator beam is typically 20 μrad, the energy resolution of the 4B-HRM is $\Delta E_{\text{4B-HRM}}$ = 0.52 meV (peak reflectivity ~ 0.5), which perfectly matches the resolution of the 4B-CDDS analyzer with the above parameters $\varphi$ = 2°, $\theta_{e1} \approx$ 1° and $\theta_{e2} \approx$ 3°. Meanwhile, the steepness of the spectral tails is similar for both the 4B-HRM and the 4B-CDDS analyzer.

The most critical components of the IXS optics in Fig. 6 are the 4B-CDDS analyzer and the collimating mirrors. The two-dimensional collimating mirrors are required because of two reasons. First, the vertical angular acceptance of the 4B-CDDS analyzer is about 0.1 mrad, much smaller than that required by IXS experiments (up to 10 mrad). Second, although the angular acceptance of 4B-CDDS along the horizontal direction is relatively broader, large horizontal divergence can cause bandwidth shifts and thus smear the energy resolution (Sturhahn & Toellner, 2011). Therefore, developing high-efficiency multilayer mirrors for collimating the scattered beam within 0.1 mrad along both directions is critical. Two-bounce L-shaped Motel multilayer mirrors (Honnicke *et al.*, 2010) have been experimentally demonstrated to be promising for this purpose. Recently, Sturhahn & Toellner (2011) also proposed the use of the 4B-HRM for both the monochromator and the analyzer. The advantage of this scheme will be that the IXS spectrometer can be designed to work at any photon energies. It should be noted that the angular acceptance of 4B-HRM is generally less than 30 μrad for achieving sufficient efficiency at $E$ ~ 10 keV, which is adequate for the monochromator. For the analyzer, however, this small acceptance will cause extremely stringent requirements for both the collimating and focusing mirrors. In comparison, here since the 4B-CDDS analyzer has a much larger acceptance (~0.1 mrad and could be further increased by using germanium as the collimator *C*), these requirements are significantly relaxed and may be more practical [see Cai (2010) for more details about the technical requirements].

## 4. Summary

We have made detailed theoretical analyses of the multiple-beam diffraction effect in the back-reflection-based CDTS and CDTDS monochromators. It is shown that this effect does exist in 008 back reflection, which causes relative efficiency losses of at least 6.3% and 12% for CDTS and CDTDS, respectively, even under the optimized diffraction conditions. These losses may not pose a fatal threat to the CDTS and CDTDS schemes, but are quite unfavorable as the maximum theoretical reflectivity values of CDTS and CDTDS are only 0.34 and 0.37, respectively. To improve the efficiency, we have proposed a 4B-CDDS scheme that completely avoids multiple-beam diffraction and transmission-induced absorption. Consequently, the optical efficiency can be improved by about 50% relative to that of CDTS or CDTDS. Meanwhile, the lengths of the *D* crystals are also significantly shortened. The 4B-CDDS analyzer, as a promising alternative scheme, may work with the 4B-HRM to perform meV- or sub-meV-resolution spectroscopy at medium photon energies.

The author is grateful to L. Young and M. Beno for encouragement and support. Thanks are also due to Y. Q. Cai, D. P. Siddons, M. G. Honnicke and L. Assoufid for helpful discussions. This work was supported by the U.S. Department of Energy, Office of Science, Office of Basic Energy Sciences, under Contract No. DE-AC02-06CH11357.

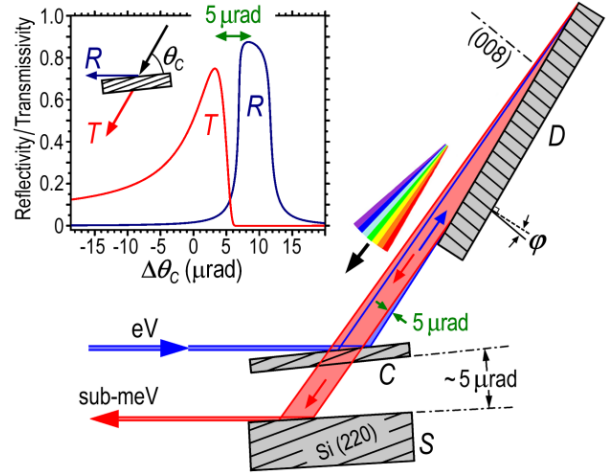

**Figure 1**

Schematic of the CDTS diffraction setup. The inset shows the Borrmann transmission curve (*T*) in comparison with the 220 Bragg reflection curve (*R*) of crystal *C* crystal in the large-incidence-small-exit geometry.

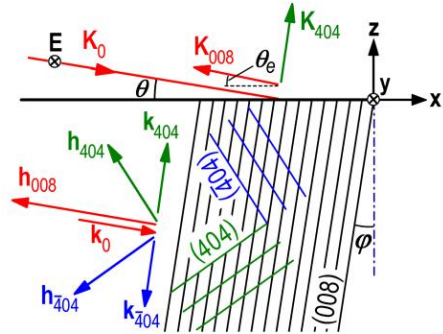

**Figure 2**

σ-polarization coplanar 4-beam diffraction configuration associated with 008 back reflection. **x**, **y** and **z** are orthogonal unit vectors.

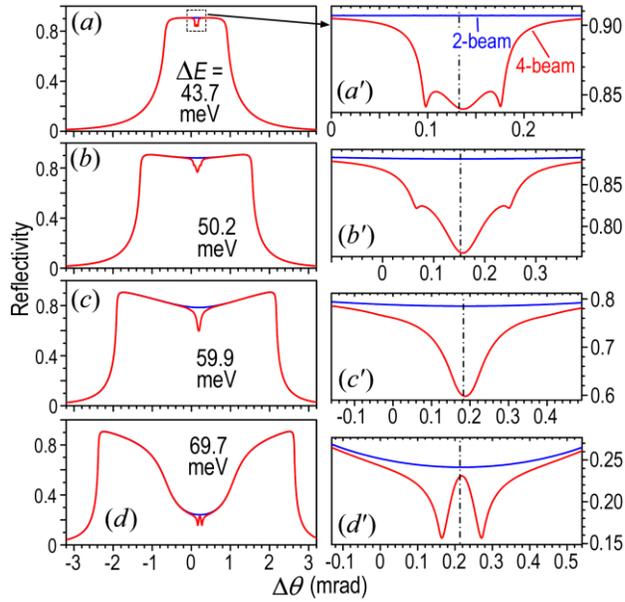

**Figure 3**

Darwin curves of 008 back reflection with $\varphi = 2°$. Each curve was calculated under a constant incident plane wave [with photon energy $E_B + \Delta E$, the values of $\Delta E$ indicated in (a)-(d)] but with the crystal rotated. The rocking angle $\Delta\theta$ is relative to $\theta_0 = \varphi = 2°$. (a')-(d') are the magnified views showing the 4-beam diffraction intensity dips in (a)-(d), respectively.

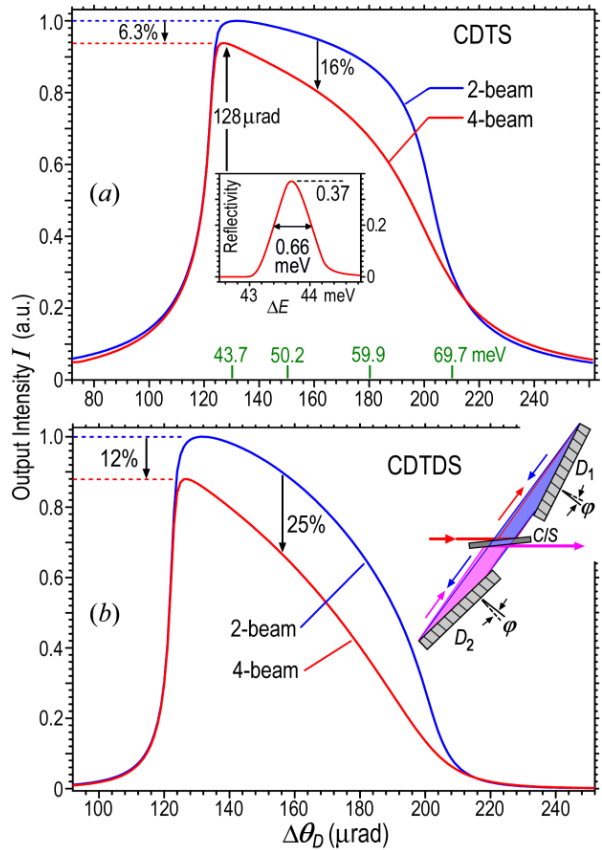

**Figure 4**

Variation of the total CDTS diffraction intensity $I$ as a function of the $D$ crystal rocking angle $\Delta\theta_D$. $\varphi = 2°$. The 2-beam curve was calculated with the parasitic 404 and $\overline{4}04$ reflections artificially ignored for crystal(s) $D$. (a) $D$ scan rocking curve of the CDTS analyzer. The inset shows the resolution function of CDTS at $\Delta\theta_D = 128$ µrad (calculated with 4-beam diffraction of crystal $D$ taken into account). (b) $D$ scan rocking curve of the CDTDS analyzer (with the diffraction setup in the inset). Crystals $D_1$ and $D_2$ are assumed to be rotated simultaneously along opposite directions ($D_1$ counterclockwise) with the same speed (with the $C/S$ crystal fixed).

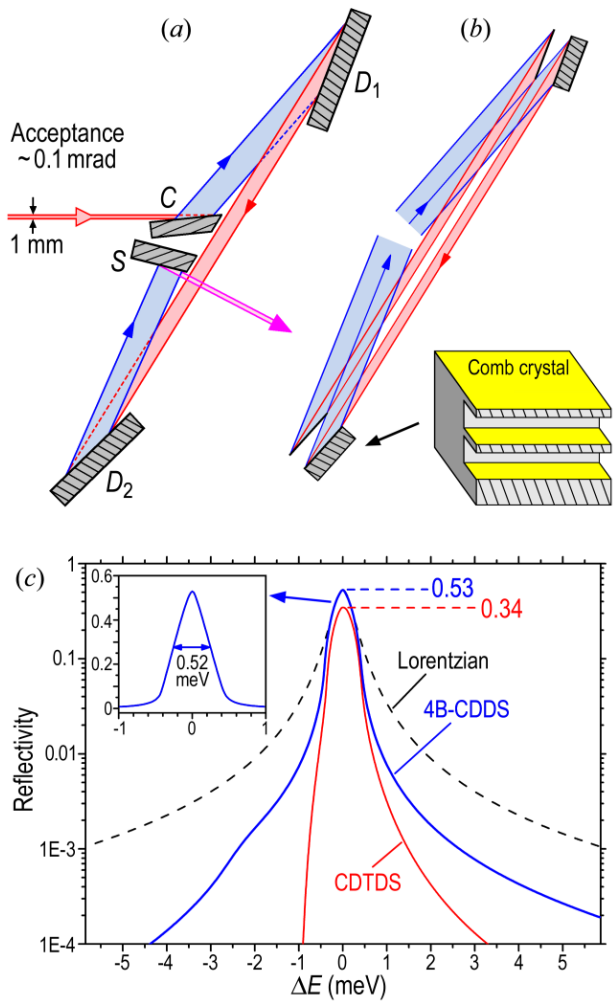

**Figure 5**
(*a*) Scheme of the 4B-CDDS analyzer. (*b*) Use of shorter 'comb crystals' as the dispersing crystals. Note that the side surfaces [nearly paralle to (008) planes] do not participate in the multi-crystal diffraction since they have a slightly different diffraction angle (Shvyd'ko, 2008). (*c*) The energy resolution function of 4B-CDDS calculated with the following parameters: 220 reflection for $C$ ($b_C = -0.0465$), $\varphi = 2°$ and $\theta_B^{008} = 89°$ for $D_1$ and $D_2$, and 440 reflection for S ($b_S = -6.7$). The inset shows the resolution function of 4B-CDDS on the linear scale. The Lorentzian distribution function (dashed line) has the same peak reflectivity and peak width as the 4B-CDDS curve. The CDTDS curve (red line) was calculated with $\varphi = 2°$ [see Fig. 4(*b*)].

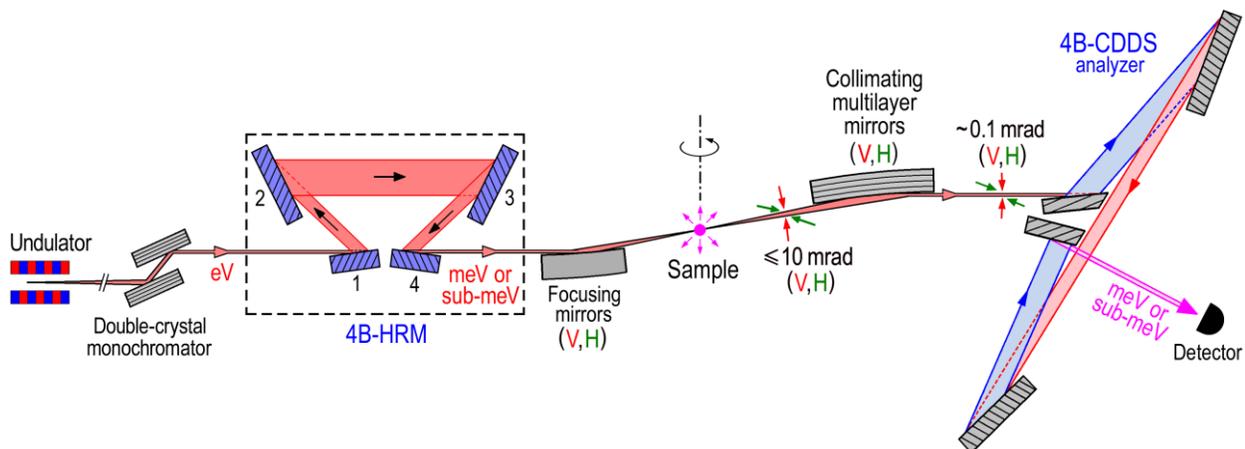

**Figure 6**
Basic components and layout of a proposed meV or sub-meV medium-energy IXS beamline. The mirrors are for two-dimensional (vertical and horizontal) focusing or collimation. Energy scan of the 4B-HRM can be realized by two-axis angular scanning of weak-link crystals 1-2 and 3-4 along opposite directions.

13